\documentclass[aps,prd,showpacs,
preprint
]{revtex4}
\usepackage{graphicx}
\usepackage{hyperref}
\usepackage{amssymb, amsmath}
\usepackage{epsf}

\def\la{\; \raise0.3ex\hbox{$<$\kern-0.75em\raise-1.1ex\hbox{$\sim$}}\;}
\def\ga{\;  \raise0.3ex\hbox{$>$\kern-0.75em\raise-1.1ex\hbox{$\sim$}}\;}

\def\pFn{p_{\raise-0.3ex\hbox{{\scriptsize F$\!$\raise-0.03ex\hbox{\rm n}}}}
}  % p_Fn
\def\pFp{p_{\raise-0.3ex\hbox{{\scriptsize F$\!$\raise-0.03ex\hbox{\rm p}}}}
}  % p_Fp
\def\pFe{p_{\raise-0.3ex\hbox{{\scriptsize F$\!$\raise-0.03ex\hbox{\rm e}}}}
}  % p_Fe
\def\pFmu{p_{\raise-0.3ex\hbox{{\scriptsize F$\!$\raise-0.03ex\hbox{
\rm $\mu$}}}} }  % p_Fe
\def\m@th{\mathsurround=0pt }
\def\eqalign#1{\null\,\vcenter{\openup1\jot \m@th
   \ialign{\strut$\displaystyle{##}$&$\displaystyle{{}##}$\hfil
   \crcr#1\crcr}}\,}
\def\dd{\mbox{d}}

\begin{document}

%%%%%%%%%%%%%%%%%%%%%%%%%%%%%%%%%%%%%%%%%%%%%%%%%%%%%%%%%%%%%%%%%%%%%%
\title{Bulk viscosity of superfluid hyperon stars}
%%%%%%%%%%%%%%%%%%%%%%%%%%%%%%%%%%%%%%%%%%%%%%%%%%%%%%%%%%%%%%%%%%%%%%
%
\author{Mikhail E. Gusakov, Elena M. Kantor}
\affiliation{
Ioffe Physical Technical Institute,
Politekhnicheskaya 26, 194021 Saint-Petersburg, Russia
%e-mail: gusakov@astro.ioffe.ru
}
\date{}
%

%\preprint{}
\pacs{
%04.40.Dg,  %Relativistic stars: structure, stability, and oscillations
%97.60.Jd,  %Neutron stars
%26.60.+c,  %Nuclear matter aspects of neutron stars in nuclear physics
%95.30.Sf   %Relativity and gravitation
97.60.Jd,
%04.30.Db,
%04.40.Dg,
26.60.+c,
47.37.+q,
47.75.+f        % Relativistic fluid dynamics
}

%%%%%%%%%%%%%%%%%%%%%%%%%%%%%%%%%%%%%%%%%%%%%%%%%%%%%%%%%%%%%%%%%%%%%%
\begin{abstract}
We calculated bulk viscosity due 
to non-equilibrium weak processes
in superfluid nucleon-hyperon matter 
of neutron stars.
For that, the dissipative relativistic hydrodynamics, 
formulated in paper \cite{gusakov07} 
for superfluid mixtures, was extended 
to the case when both nucleons and hyperons are superfluid.
It was demonstrated that in the most general case 
(when neutrons, protons, $\Lambda$, and $\Sigma^{-}$ hyperons 
are superfluid), 
non-equilibrium weak processes generate 
{\it sixteen} bulk viscosity coefficients,
with only {\it three} of them being independent.
In addition, we corrected an inaccuracy
in a widely used formula for 
%ordinary 
the bulk viscosity
of non-superfluid nucleon-hyperon matter.
\end{abstract}
%%%%%%%%%%%%%%%%%%%%%%%%%%%%%%%%%%%%%%%%%%%%%%%%%%%%%%%%%%%%%%%%%%%%%%

\maketitle

%%%%%%%%%%%%%%%%%%%%%%%%%%%%%%%%%%%%%%%%%%%%%%%%%%%%%%%%%%%%%%%%%%%%%%%%%%%%%
%**************** Section 1 ******************************
\section{Introduction}
\label{1}
%%%%%%%%%%%%%%%%%%%%%%%%%%%%%%%%%%%%%%%%%%%%%%%%%%%%%%%%%%%%%
It is well known (see, e.g.,
\cite{Lindblom95,ak01,andersson03,andersson06}), 
that neutron stars can be unstable 
with respect to the emission of gravitational waves.
The matter in a pulsating neutron star 
is not (even locally) in chemical equilibrium.
The relaxation towards chemical equilibrium is accompanied 
by the dissipation of pulsation energy.
This process is one of the most important dissipative processes, 
suppressing the growth of gravitational-wave instability.
It can be described by the introduction of an
effective bulk viscosity 
in the hydrodynamic equations
(see, e.g., \cite{ll87}).

The bulk viscosity due to non-equilibrium weak processes
in neutron stars was calculated 
by a number of authors
(for a review see, e.g., \cite{dsw07}).
It strongly depends on the composition of stellar matter. 
In this paper we consider the matter of inner layers of neutron stars, 
composed of electrons, muons, neutrons, protons, 
$\Lambda$, and $\Sigma^{-}$ hyperons (nucleon-hyperon matter).
A calculation of the bulk viscosity for nucleon-hyperon matter
is complicated by the fact that baryons 
in such matter can be superfluid
\cite{yls99,ls01,yp04,bb98}.

The bulk viscosity of superfluid matter was calculated 
in a number of papers (see, e.g., \cite{hly00,hly01,hly02,lo02,no06,ssr06}).
In these papers only one bulk viscosity coefficient was studied, 
analogous to that in non-superfluid hydrodynamics.
The effects of superfluidity were taken into account 
only in calculating the reaction rates.
However, it is known that there are several bulk viscosity
coefficients in hydrodynamics of superfluid liquid
\cite{putterman74,ll87,khalatnikov89,gusakov07}.

In a recent paper \cite{gusakov07} 
{\it four} bulk viscosity coefficients were calculated 
for the matter composed of superfluid neutrons, 
superfluid protons and electrons.
It was shown that taking into account 
three additional bulk viscosity coefficients
results in a significant decrease of characteristic damping times 
of sound modes (approximately, by a factor of three).

In this paper we extend the results of Ref. \cite{gusakov07} 
to the case of nucleon-hyperon matter. 
In particular, we demonstrate 
how the dissipative 
hydrodynamics \cite{gusakov07} 
should be modified to describe a
possible presence of superfluid hyperons.
Next, we show that 
in the most general case, 
when baryons of any species are superfluid,
non-equilibrium processes of mutual transformations of particles
generate {\it sixteen} bulk viscosity coefficients, 
only three of them being independent.
In addition, we correct an inaccuracy 
in the expression for the bulk viscosity of non-superfluid 
nucleon-hyperon matter made in Ref. \cite{lo02}
and spreaded widely in the literature 
(see, e.g., \cite{vanD04,drago05,mohit06,debarati06,pan06,debarati07}). 
We calculate the bulk viscosity correctly
and compare our results with those of Ref. \cite{lo02}.

The paper is organized as follows.
In section II we calculate the bulk viscosity of non-superfluid 
nucleon-hyperon matter.
In section III we calculate and analyze 
all {\it sixteen} bulk viscosity coefficients 
describing dissipation 
in superfluid nucleon-hyperon mixtures; 
the relations between these 
coefficients are also discussed.
Section IV presents a summary.

%%%%%%%%%%%%%%%%%%%%%%%%%%%%%%%%%%%%%%%%%%%%%%%%%%%%%%%%%%%%%%%%%%%%%%%
\section{Bulk viscosity of non-superfluid hyperon matter}
\label{2}
%%%%%%%%%%%%%%%%%%%%%%%%%%%%%%%%%%%%%%%%%%%%%%%%%%%%%%%%%%%%%%%%%%%%%%%

In this section we derive 
an expression for the bulk viscosity 
due to non-equilibrium processes 
of particle transformations 
in a dense non-superfluid matter 
composed of electrons ($e$), muons ($\mu$),
neutrons ($n$), protons ($p$), and hyperons
($\Lambda$ and $\Sigma^{-}$ hyperons).
Here and below the variation $\delta A$
of some physical quantity $A$ will be defined as
the difference $A-A_0$,
where $A_0$ is the value of $A$
in thermodynamic equilibrium
(when matter is unperturbed).

The most effective weak processes in nucleon-hyperon matter
are the following non-leptonic reactions
\cite{Jones01,hly02,lo02,schaff08}
\begin{eqnarray}
n+n &\leftrightarrow& p + \Sigma^-,
\label{s}\\
n+p &\leftrightarrow& p +\Lambda,
\label{l}\\
n+n &\leftrightarrow& n+ \Lambda,
\label{ln}\\
n+\Lambda &\leftrightarrow& \Lambda+ \Lambda.
\label{ll}
\end{eqnarray}
The full thermodynamic equilibrium 
implies the equilibrium with respect 
to these reactions,
\begin{eqnarray}
&&2 \mu_{n0} - \mu_{p 0} - \mu_{\Sigma 0} =0,
\label{s_eq} \\
&& \mu_{n0}-\mu_{\Lambda 0}=0.
\label{l_eq}
\end{eqnarray}
Here $\mu_{i0}$ are the chemical potentials 
of particle species
$i=n,p,\Lambda,\Sigma$ 
taken at equilibrium 
(correspondingly, $\mu_{i}$ are the chemical potentials 
in the perturbed matter).
Notice that, the equilibrium conditions for reactions
(\ref{l}), (\ref{ln}), and (\ref{ll}) coincide.

Leptonic reactions 
(e.g, direct and modified Urca processes 
with electrons or muons) 
are much slower in comparison to 
the reactions (\ref{s})--(\ref{ll}). 
For `typical' perturbation frequencies
(e.g., $10^3-10^4\; s^{-1}$ for radial modes 
or for r-modes of rapidly rotating neutron stars)
the leptonic reactions cannot influence substantially
the chemical composition of perturbed matter. 
Hence, the main contribution to the 
bulk viscosity comes from the non-leptonic 
reactions (\ref{s})--(\ref{ll}).
In addition to the processes described above, 
there is a fast non-leptonic reaction 
due to the strong interaction of baryons 
\begin{equation}
n+\Lambda \leftrightarrow p + \Sigma^-.
\label{fast}
\end{equation}
In accordance with Ref.\ \cite{lo02} 
we assume that the perturbed matter is always 
in equilibrium with respect to this reaction,
\begin{equation}
\delta \mu_{\rm fast} \equiv
\mu_{n}+\mu_\Lambda-\mu_{p}-\mu_\Sigma
%=\delta \mu_{n} + \delta \mu_\Lambda
%-\delta \mu_{p} - \delta \mu_\Sigma 
=0.
\label{eqfast}
\end{equation}
Let us obtain the expression 
for the bulk viscosity of non-superfluid 
nucleon-hyperon matter. 
For that, we consider a pulsating nucleon-hyperon matter, 
slightly perturbed from an equilibrium state 
(so that one can use the linear perturbation theory).
%(with the pressure $P_0$).
%Without any loss of generality
%we can assume that the perturbation 
%is periodic in time, so all the thermodynamic 
%functions oscillate near 
%their equilibrium values 
%with the frequency $\omega$.
If the reactions (\ref{s})--(\ref{ll}) 
are forbidden, then pulsations are reversible 
and there is no energy dissipation 
(notice that, 
%at the same time 
the reaction \ref{fast} is open).
We denote the pressure in this case by $P_{\rm eq}$. 
The presence of the reactions (\ref{s})--(\ref{ll}) 
in the pulsating matter leads to a difference 
between the real pressure $P$ and $P_{\rm eq}$. 
We define the bulk viscosity $\xi$ by the formula
\begin{equation}
P-P_{\rm eq} \equiv -\xi \,
{\rm div} (\pmb{u}),
\label{visk}
\end{equation}
where $\pmb{u}$ is the hydrodynamic 
velocity of pulsations. 
Notice that, this definition
differs from the usually accepted one 
(see, e.g., Ref. \cite{lo02}). 
Usually, instead of $P_{\rm eq}$ 
in formula (\ref{visk}) it is common to substitute the pressure
which would be established in the pulsating matter assuming 
that there is an
equilibrium with respect to all the reactions 
(i.e. the reactions are very fast).
Both these approaches are possible.

Generally, the pressure $P$
and the other thermodynamic quantities
%are the functions of 
depend on 
six parameters, 
for example, the number densities $n_j$, 
where $j=n, p, \Lambda, \Sigma, e, \mu$ 
(one can neglect the dependence on temperature 
in strongly degenerate neutron-star matter, 
see, e.g., Refs.\ \cite{reisenegger95,gyg05,gusakov07}). 
However, these parameters are not all independent
because in the nucleon-hyperon matter two conditions 
should be satisfied: the equilibrium condition
(\ref{eqfast}) with respect to the reaction (\ref{fast}) 
and the condition of quasineutrality,
\begin{equation}
n_{p}=n_{e}+n_\mu + n_\Sigma.
\label{charge neutrality}
\end{equation}
Taking into account that the reactions (\ref{s})--(\ref{ll}) 
and (\ref{fast}) conserve the number of leptons
and the leptonic processes are neglected,
we obtain that the relative number densities 
of leptons $x_e \equiv n_e/n_b$ and
$x_\mu \equiv n_{\mu}/n_b$ 
($n_{\rm b} \equiv n_n+n_p+n_\Lambda+n_\Sigma$ 
is the baryon number density)
remain constant during the pulsations, 
\begin{equation}
\delta x_e=\delta x_{\mu}=0. 
\label{dxedxm}
\end{equation}
%
%$\delta x_e=\delta x_{\mu}=0$. 
%
This result is valid only for non-superfluid matter and follows from the 
continuity equations for baryons, electrons, and muons,
\begin{eqnarray}
\frac{\partial\delta n_{b}}{\partial t}+{\rm div}(n_{b} \pmb{u}) &=&
0, \label{nb}\\
\frac{\partial\delta n_{e}}{\partial t}+{\rm div}(n_{e}
\pmb{u}) &=& 0,
\label{ne} \\
\frac{\partial\delta n_\mu}{\partial t}
+{\rm div}(n_\mu \pmb{u}) &=& 0.
\label{nmu}
\end{eqnarray}
%
%Here $n_{\rm b} \equiv n_n+n_p+n_\Lambda+n_\Sigma$ 
%is the baryon number density.
If baryons of any species $n$, $p$, $\Lambda$ and/or $\Sigma$ 
are superfluid, then 
the continuity equation for baryons (\ref{nb}) should be modified (see Section 3)
and Eq. (\ref{dxedxm}) does not hold.

%Taking into account Eq.\ (\ref{dxedxm})
%the quasineutrality condition (\ref{charge neutrality}) 
%can be rewritten in the form 
%(see also the derivation 
%of the formula 3.18 in Ref.\ \cite{lo02})
%
%\begin{equation}
%\delta x_{p}=\delta x_\Sigma,
%\label{charge neutrality1}
%\end{equation}
%
%where $x_p \equiv n_p/n_b$;
%$x_\Sigma \equiv n_\Sigma/n_b$.
%%$n_b=n_n+n_p+n_\Lambda+n_\Sigma$
%%is the number density of baryons.

In view of Eqs.\ (\ref{eqfast}) and (\ref{charge neutrality}), 
%(or, equivalently, Eq.\ (\ref{charge neutrality1}))
pressure is a function 
of only four independent variables, 
say, $n_{b}, n_{H}, x_{e}$, and $x_{\mu}$ 
($n_{H} \equiv n_\Lambda+n_\Sigma$ is the hyperon number density).
%We will choose $n_{b}, n_{H}, n_{e}$, and $n_{\mu}$ 
%as these variables, 
%where $n_{H}=n_\Lambda+n_\Sigma$ 
%is the hyperon number density. 
Expanding $P(n_b,n_H,x_e,x_{\mu})$ in the Taylor series 
near the equilibrium state, 
one obtains for the variation of pressure $\delta P$,
\begin{equation}
\delta P=\frac{\partial P(n_{b},n_{H}, x_{e}, x_{\mu})}{\partial n_{b}} \delta n_{b}
+\frac{\partial P(n_{b},n_{H}, x_{e}, x_{\mu})}{\partial n_{H}} \delta n_{H},
%+\frac{\partial P}{\partial n_{e}}
%\delta n_{e}+\frac{\partial P}{\partial n_\mu}
%\delta n_\mu.
\label{dP}
\end{equation}
where we used Eq.\ (\ref{dxedxm}).
The variations $\delta n_{b}$ and $\delta n_{H}$ can be found from 
the continuity equations for baryons (\ref{nb}) and hyperons,
\begin{equation}
\frac{\partial\delta n_{H}}{\partial t}+{\rm div}(n_{H}
\pmb{u}) = \Delta \Gamma_1 + \Delta \Gamma_2 + \Delta \Gamma_3 + \Delta 
\Gamma_4.
%\lambda_1 \delta\mu_1+ \lambda_2 \delta\mu_2+
%\lambda_3 \delta\mu_3,
\label{nH}
\end{equation}
Here $\Delta \Gamma_1$, $\Delta \Gamma_2$, $\Delta \Gamma_3$, and $\Delta 
\Gamma_4$
are the net numbers of hyperons generated in a unit volume per unit time 
in reactions (\ref{s}), (\ref{l}), (\ref{ln}), and (\ref{ll}), respectively.
If deviation from the equilibrium state is small,
the sources
$\Delta \Gamma_l$ ($l$=1,$\ldots$,4)
can be expressed as
(see, e.g, \cite{hly00,hly01,hly02})
\begin{equation}
\Delta \Gamma_l=\lambda_l \delta \mu_l,
\label{source}
\end{equation}
where $\lambda_l$ are the `reaction rates',
some 
%certain 
functions of number densities and temperature;
$\delta\mu_1 \equiv 2\mu_{n}-\mu_{p}-\mu_\Sigma$,
$\delta\mu_2 = \delta\mu_3 = \delta\mu_4 \equiv \mu_{n}-\mu_\Lambda$
are the chemical potential disbalances for 
the reactions (\ref{s}), (\ref{l}), (\ref{ln}), and (\ref{ll}), respectively.
Taking into account the equilibrium condition (\ref{eqfast})
for the fast reaction (\ref{fast}), one has:
$\delta\mu_1=\delta\mu_2=
\delta\mu_3=
\delta\mu_4 \equiv \delta\mu$.

Notice that, there is no source in the equation (\ref{nH})
owing to the fast reaction (\ref{fast}), 
because this reaction does not change the number of hyperons. 
The choice of another variable instead of $n_{H}$
(for example, the neutron number density $n_{n}$) 
would make it necessary to take into account 
the source due to the reaction (\ref{fast}). 
In the paper of Lindblom and Owen \cite{lo02} 
(and in the subsequent papers 
\cite{vanD04,drago05,mohit06,debarati06,pan06,debarati07}) 
the number density of neutrons was chosen as such variable, 
but the source of neutrons owing to the fast reaction 
(\ref{fast}) was neglected.
This leads to an error in the expression for the bulk viscosity.

In fact, one could think
that the source of neutrons due 
to the fast reaction (\ref{fast}) equals zero, 
because the matter, as we mentioned before, 
is in equilibrium with respect to this reaction. 
However, this is not quite true, because even small 
(negligible in all other situations) 
deviation from the equilibrium $\delta\mu_{\rm fast}
=\mu_{n}+\mu_\Lambda-\mu_{p}-\mu_\Sigma$ 
multiplied by the large reaction rate 
$\lambda_{\rm fast}$ results 
in a finite (non-zero) source $\Delta \Gamma_{\rm fast}
= \lambda_{\rm fast} \delta\mu_{\rm fast}$. 
This fact was emphasized
by Jones \cite{Jones01}.

%Without any loss of generality
%we can assume 
Now let us assume that the perturbation of matter
is periodic, so that all the thermodynamic 
quantities oscillate near 
their equilibrium values 
with the frequency $\omega$.
Then one finds from the continuity equations 
(\ref{nb}) and (\ref{nH})
%Taking into account that the perturbation 
%of matter depends on time as 
%${\rm exp}(i \omega t)$, 
%we find from the continuity equations 
%
\begin{eqnarray}
\delta n_{b}=-\frac{n_{b}}{i\omega}{\rm div}(\pmb{u}),
\label{dni begin}\\
\delta n_{H}=-\frac{1}{i\omega}[n_{H}
{\rm div}(\pmb{u})-\lambda \, \delta\mu], 
\label{dnH}
\end{eqnarray}
where $\lambda \equiv \lambda_1+\lambda_2+\lambda_3+\lambda_4$.
As in the case of the pressure, the chemical potential disbalance 
$\delta \mu$ in Eq.\ (\ref{dnH})
is a function of $n_b$, $n_H$, $x_e$, and $x_{\mu}$.
In analogy with Eq.\ (\ref{dP}) for $\delta P$,
it can be expanded near the equilibrium state and written as
\begin{equation}
\delta \mu =\frac{\partial \delta \mu(n_b,n_H,x_e,x_{\mu})}{\partial n_{b}}
\delta n_{b}+\frac{\partial \delta \mu(n_b,n_H,x_e,x_{\mu})}{\partial n_{H}}
\delta n_{H}.
\label{dmu1}
\end{equation}
Here we take into account that $\delta \mu =0$ 
in the full equilibrium and that the relative number densities 
$x_{e}$ and $x_{\mu}$ do not change 
in the course of pulsations (see Eq.\ (\ref{dxedxm})).

Now, solving the system of equations 
(\ref{dni begin})--(\ref{dmu1}) and substituting the 
expressions for $\delta n_b$ and $\delta n_H$
into Eq. (\ref{dP}), we derive
\begin{eqnarray}
P-P_{\rm eq}&=& {\rm div}(\pmb{u})  \,
\frac{\partial P(n_b,x_H,x_e,x_{\mu})}{\partial
x_{H}} \, \frac{\lambda }{\omega^2} \, \frac{\partial 
\delta\mu(n_b,x_H,x_e,x_{\mu})}{\partial n_{b}} 
\nonumber \\
&& \times \left[\frac{i \lambda}{\omega n_{b}}
\frac{\partial \delta \mu(n_b,x_H,x_e,x_{\mu})}{\partial x_{H}} +1 \right]^{-1}.
\label{PminPeq}
\end{eqnarray}
Here the independent variables are 
$n_b$, $x_H \equiv n_H/n_b$, $x_e$, and $x_{\mu}$.
It is easy to express the bulk viscosity $\xi$ from this equation. 
We are mainly interested in the real part of $\xi$ because it is
${\rm Re}(\xi)$ that is responsible for the energy dissipation 
(see, e.g., \cite{hly00}).
In this sense it is probably more appropriate to define
${\rm Re}(\xi)$ as the `real' bulk viscosity.
It equals
\begin{equation}
{\rm Re}\;\xi= -\frac{n_{b}^2}{\lambda} \, \frac{\partial
P(n_b,x_H,x_e,x_{\mu})}{\partial x_{H}} \, \frac{\partial 
\delta \mu(n_b,x_H,x_e,x_{\mu})}{\partial n_{b}}
\left[ \frac{\partial \delta \mu(n_b,x_H,x_e,x_{\mu})}{\partial
x_{H}}\right]^{-2} \frac{1}{1+\omega^2 \tau^2},
\label{ksi}
\end{equation}
where $\tau \equiv n_{b}/\lambda \;
\left[\partial \delta \mu(n_b,x_H,x_e,x_{\mu})/\partial x_{H}\right]^{-1}$.
%In Eqs.\ (\ref{PminPeq}) and (\ref{ksi})
%the independent variables are 
%$n_b$ and $x_H \equiv n_{H}/n_{b}$.

For comparison, we present here the result
of Lindblom and Owen \cite{lo02} 
(notice that, these authors neglected
the reactions \ref{ln} and \ref{ll}, 
thus assuming that $\lambda_3=\lambda_4=0$)
\begin{equation}
{\rm Re}\;\xi_{\rm L}= -\frac{n_{b}^2}{2\lambda_1+\lambda_2} \,
\frac{\partial P(n_b,x_n,x_e,x_{\mu})}{\partial x_{n}} \,
\frac{\partial \delta \mu(n_b,x_n,x_e,x_{\mu})}{\partial n_{b}}
\left[ \frac{\partial \delta \mu(n_b,x_n,x_e,x_{\mu})}{\partial
x_{n}}\right]^{-2} \frac{1}{1+\omega^2 \tau_{L}^2},
\label{ksi_L}
\end{equation}
where $\tau_{\rm L} \equiv n_{b}/(2\lambda_1+\lambda_2) \;
\left[\partial \delta \mu(n_b,x_n,x_e,x_{\mu})/\partial x_{n}\right]^{-1}$; 
$x_n \equiv n_n/n_b$.

The bulk viscosity (\ref{ksi}) 
depends on the reaction rates 
$\lambda_1,
\lambda_2$, $\lambda_3$, and $\lambda_4$.
Some of them were calculated in a number of papers
\cite{Jones01,lo02,hly02,vanD04,schaff08}.
Different authors present different results for the rates;
the discussion of advantages and disadvantages 
of their calculations can also be found in those papers.
Unfortunately, to date, there are no strict calculations 
of the reaction rates. 
It is reasonable to think that 
all the rates are of the same order of magnitude.

The dependence of the bulk viscosity on the baryon number density 
for the temperature $T=3\times10^9$ K 
and the oscillation frequency $\omega=10^4$ s$^{-1}$ 
is presented in Fig.\ 1. 
While calculating the bulk viscosity, 
we used the third equation of state 
of Glendenning \cite{g85}.

%%%%%%%%%%%%%%%%%%%%%%%%%%%%%%%%%%%%%%%%%%%%%%%%%%%%%%%%%%%%%%%%%%%%
\begin{figure}
\begin{center}
\leavevmode \epsfxsize=8cm \epsfbox[60 200 560 670]{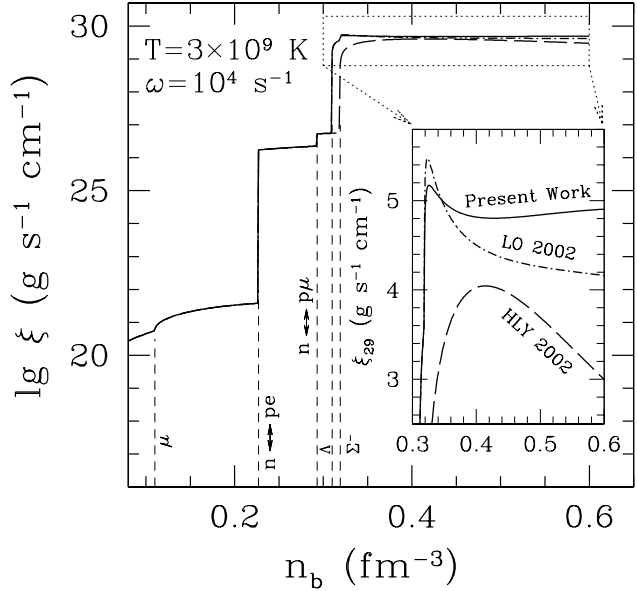}
\end{center}
\caption{ Bulk viscosity $\xi$ versus baryon number density
$n_{b}$ at $T=3\times10^9$ K and $\omega=10^4$ s$^{-1}$ 
for non-superfluid matter. 
Solid, long-dashed, and dot-dashed lines show our results, 
the results of Ref.\ \cite{hly02} and Ref.\ \cite{lo02}, respectively.
Vertical dashes indicate the thresholds for
(from left to right):
appearance of muons; direct Urca processes 
involving electrons and muons, respectively;
appearance of $\Lambda$ and $\Sigma^{-}$ hyperons, respectively.
The inset demonstrates the difference 
between our calculations and those 
of Refs.\ \cite{hly02} and \cite{lo02} in more detail.
}
\label{fig:fig1}
\end{figure}
%%%%%%%%%%%%%%%%%%%%%%%%%%%%%%%%%%%%%%%%%%%%%%%%%%%%%%%%%%%%%%%%%%%%
The solid line illustrates {\it our} results for the bulk viscosity obtained
from Eq.\ (\ref{ksi}).
We employed the reaction rates from Ref.\ \cite{lo02} 
and, following that paper, we put the rates of the reactions 
(\ref{ln}) and (\ref{ll}) equal zero, $\lambda_{3}=\lambda_{4}=0$.
The dot-dashed line is the bulk viscosity calculated 
as described in Ref.\ \cite{lo02} 
(see also formula \ref{ksi_L}).
%We remind that in the paper \cite{lo02} the baryon 
%number density $n_{b}$ and 
%the relative number density of 
%neutrons $x_{n}$ were chosen as independent 
%variables.
We remind that in that paper the relative number density of 
neutrons $x_{n}$ was chosen as one of the independent variables. 
%At the same time 
However,
the source of neutrons due to the fast reaction (\ref{fast})
was erroneously neglected.
As one can see, this mistake does not influence the results significantly 
(typically, by $\sim$ (10 -- 30) \%).
By the long dashes we show the results of Ref.\ \cite{hly02}.
In Ref.\ \cite{hly02}, only one hyperon reaction 
was taken into account, namely the reaction (\ref{s}).
As in Ref.\ \cite{lo02}, the source due to the fast reaction 
(\ref{fast}) was neglected 
and, in addition, the equilibrium condition (\ref{eqfast})
with respect to this reaction was ignored. 
Moreover, the authors of Ref.\ \cite{hly02} 
used the non-relativistic approximation
when calculating the rate of the reaction (\ref{s}).
This assumption is not well justified for 
the baryon number densities 
in the range $n_b \ga (0.3-0.6)$ fm$^{-3}$
because of the strong dependence of 
the reaction rates on $n_b$ (see, e.g., Ref.\ \cite{lo02}).

%%%%%%%%%%%%%%%%%%%%%%%%%%%%%%%%%%%%%%%%%%%%%%%%%%%%%%%%%%%%%%%%%
\section{Bulk viscosity of superfluid nucleon-hyperon matter}
%%%%%%%%%%%%%%%%%%%%%%%%%%%%%%%%%%%%%%%%%%%%%%%%%%%%%%%%%%%%%%%%%

In this section, unless otherwise is stated, 
the subscripts $i$ and $k$ refer 
to baryons ($i,k = n, p, \Lambda, \Sigma$). 
The summation is assumed over
repeated baryon indices $i$~and~$k$. 
The subscript $l$ refers to leptons 
($l=e, \mu$); the subscript $j$ 
runs over all particle species
($j=n, p, \Lambda, \Sigma, e, \mu$).

%%%%%%%%%%%%%%%%%%%%%%%%%%%%%%%%%%%%%%%%%%%%%%%
\subsection{The relativistic hydrodynamics of
superfluid nucleon-hyperon mixture}
%%%%%%%%%%%%%%%%%%%%%%%%%%%%%%%%%%%%%%%%%%%%%%%
In Ref.\ \cite{gusakov07} the dissipative relativistic hydrodynamics of 
superfluid mixtures was formulated for $npe$ matter. 
Here we extend this hydrodynamics 
to the case of superfluid nucleon-hyperon matter composed of 
superfluid protons, neutrons, $\Lambda$ and $\Sigma^-$ hyperons,
as well as normal electrons and muons.

The general formulae (Eqs.\ 26--34 of Ref.\ \cite{gusakov07}) 
describing the relativistic hydrodynamics 
of superfluid mixture remain valid with the notion that 
now the subscripts $i$ and $k$ refer not only 
to superfluid nucleons ($i,k = n, p$)
but also to superfluid hyperons ($i,k = n, p, \Lambda, \Sigma$).
For instance, the continuity equations for particle species $j$ 
are written as
\begin{equation}
\partial_{\mu} j^{\mu}_{ (j) } = 0,
\label{particle_conservation}
\end{equation}
with
\begin{equation}
j^{\mu}_{(i)} = n_i u^{\mu} + Y_{ik} w^{\mu}_{(k)}, \quad
j^{\mu}_{({l})} = n_{l} u^{\mu}.
\label{particle_conservation2}
\end{equation}
Energy-momentum conservation law has the form
\begin{equation}
\partial_{\mu} T^{\mu \nu} =0,
\label{energymomentum}
\end{equation}
where
\begin{eqnarray}
&& T^{\mu \nu} = (P+\varepsilon) \, u^{\mu} u^{\nu}
+ P \eta^{\mu \nu}
\nonumber \\
&&+ Y_{ik} \left[ w^{\mu}_{(i)} w^{\nu}_{(k)}
+ \mu_i \, w^{\mu}_{(k)} u^{\nu}
+ \mu_k \, w^{\nu}_{(i)} u^{\mu} \right] + \tau^{\mu \nu}
\label{Tmunu2}
\end{eqnarray}
and $\tau^{\mu \nu}$ is the dissipative correction 
to the energy-momentum tensor which will be specified below.
It satisfies the constraint
\begin{equation}
u_{\mu} u_{\nu} \tau^{\mu \nu} =0.
\label{tau}
\end{equation}
The hydrodynamic equations 
must be supplemented by the second law of thermodynamics
\begin{equation}
\dd \varepsilon = T \, \dd S + 
\mu_i \, \dd n_i + \mu_e \, \dd n_e +\mu_{\mu} \, \dd n_{\mu}+ { Y_{ik} \over 2} \, 
\dd \left[ w^{\mu}_{(i)} w_{(k) \mu} \right].
\label{2ndlaw2} 
\end{equation}
Using the quasineutrality condition (\ref{charge neutrality})
and the condition of equilibrium (\ref{eqfast}) 
with respect to the fast reaction (\ref{fast}), 
Eq.\ (\ref{2ndlaw2}) can be rewritten as
\begin{equation}
\dd \varepsilon = T \, \dd S 
+ \mu_{n} \, \dd n_b - \delta \mu \, \dd n_H
-(\mu_{n}-\mu_p-\mu_e) \, \dd n_e -
(\mu_{n}-\mu_p-\mu_{\mu}) \, \dd n_{\mu}
+ { Y_{ik} \over 2} \, 
\dd \left[ w^{\mu}_{(i)} w_{(k) \mu} \right],
\label{2ndlaw3} 
\end{equation}
where we remind the notation $\delta \mu \equiv \mu_n-\mu_{\lambda}$.
In full thermodynamic equilibrium the third, fourth, and the fifth terms
are zero because of Eqs.\ (\ref{s_eq}) and (\ref{l_eq}), and
of the beta-equilibrium conditions, $\mu_n=\mu_p+\mu_e$ and $\mu_n=\mu_p+\mu_{\mu}$ 
(see, e.g., Refs.\ \cite{hly00,hly01}). 

In Eqs.\ (\ref{particle_conservation})--(\ref{2ndlaw3}) 
$Y_{ik}$ is a $4 \times 4$ symmetric matrix 
which is related in the non-relativistic limit to the 
entrainment matrix $\rho_{ik}$ 
%(see Refs. \cite{ab75, bjk96, gh05, ga06})
by the equality \cite{ga06, gusakov07} $Y_{ik}=\rho_{ik}/(m_i m_k)$, 
where $m_i$ is the mass of a free baryon of species $i$ 
(the matrix $\rho_{ik}$ is a natural generalization of
the superfluid density to the case of superfluid mixtures, 
see, e.g., Refs. \cite{ab75, bjk96, gh05}).
To the best of our knowledge, the matrix $Y_{ik}$ has not 
been calculated for a nucleon-hyperon matter.
Furthermore, $\varepsilon$ is the energy density; 
$\mu_j$ is the relativistic chemical potential 
of particle species $j$;
and $S$ is the entropy density. 
The pressure $P$ in Eq.\ (\ref{Tmunu2}) is defined in the same way 
as for ordinary (non-superfluid) matter 
\cite{ga06, gusakov07}, 
\begin{equation}
P=-\varepsilon + \mu_i n_i + \mu_e n_e + \mu_{\mu} n_{\mu} + T S.
\label{Pressure}
\end{equation}
Next, $\eta^{\mu \nu}={\rm diag}(-1,+1,+1,+1)$ 
in Eq.\ (\ref{Tmunu2}) is the special relativistic metric;
$u^\mu$ is the four-velocity of normal 
(non-superfluid) liquid component normalized 
so that $u_{\mu} u^{\mu}=-1$ 
(we assume that all non-superfluid components 
move with the same velocity $u^{\mu}$).
%$\mu_i$ is the chemical potential of particle species $i$.
The four-vectors $w^{\mu}_{(i)}$ 
satisfy the condition
\begin{equation}
u_{\mu} w^{\mu}_{(i)}=0
\label{uw}
\end{equation}
and describe motion of superfluid components.
To take into account the potentiality of superfluid motion, 
a four-vector $w_{(i)}^{\mu}$ 
should be expressed through some scalar functions $\phi_i$ 
and written as (see Ref.\ \cite{gusakov07})
\begin{equation}
w^{\mu}_{(i)} = \partial^{\mu} \phi_i 
- q_i A^{\mu} - (\mu_i + \varkappa_i) u^{\mu}.
\label{w_i}
\end{equation}
Here the scalar $\phi_i$ is related to the wave function phase of the 
Cooper-pair condensate $\Phi_i$ by the equality
${\pmb \triangledown} \phi_i=\hbar{\pmb \triangledown\Phi_i}/2$; 
$A^\mu$ is the four-potential of the
electromagnetic field; 
$q_i$ is the electric charge of particle species $i$; 
%$\varkappa_i$ are the small corrections appearing 
%in the hydrodynamic equations when one takes into account dissipation.
$\varkappa_i$ is a small dissipative correction to be determined below.

Note that one can avoid the introduction of new functions $\phi_i$
in the hydrodynamics of superfluid mixtures if 
one formulates the potentiality condition (\ref{w_i}) 
in the equivalent way
\begin{eqnarray}
&&\partial^{\nu} \left[ w^{\mu}_{(i)} 
+q_i A^{\mu} + (\mu_i + \varkappa_i) u^{\mu} \right]
\nonumber \\
&&= \partial^{\mu} \left[ w^{\nu}_{(i)} 
 +q_i A^{\nu} +(\mu_i + \varkappa_i) u^{\nu} \right].
\label{w_i2}
\end{eqnarray}
Below we will use the latter formulation 
because it is more suitable for our purpose.
In this approach, four-vectors $w^{\mu}_{(i)}$  
are treated as independent hydrodynamic variables. 

The hydrodynamics discussed above would be incomplete without 
%must be supplemented by the 
an indication what we mean by a {\it comoving frame}, 
that is the frame where we measure (and define) 
all the thermodynamic quantities.
%The thermodynamic quantities entering the hydrodynamics discussed above
%should be defined in some frame (the so-called {\it comoving frame}). 
%As follows from the condition (\ref{uw}) 
As was demonstrated in Ref.\ \cite{gusakov07}, 
the condition (\ref{uw}) dictates that
the comoving is the frame where
the four-velocity $u^{\mu}$ equals $u^{\mu}=(1,0,0,0)$. 
In this frame, the basic thermodynamic quantities 
$\varepsilon$, $n_j$, and ${\pmb w}_{(i)}$ 
(or ${\pmb \triangledown \phi_i}$) are defined by 
(see Eqs.\ (\ref{particle_conservation2}), 
(\ref{Tmunu2}), (\ref{tau}), and (\ref{uw}))
\begin{eqnarray}
j_{j}^0 &=& n_j, \quad \quad 
\label{condition1} \\
{\pmb j}_{i} &=& Y_{ik} \, {\pmb w}_{(k)} 
= Y_{ik} \, {\pmb \triangledown} \phi_k, \quad \quad 
\label{condition2} \\
T^{00} &=& \varepsilon.
\label{condition3}
\end{eqnarray}
All other thermodynamic quantities can be considered as their functions 
%of $\varepsilon$, $n_j$, and ${\pmb \triangledown \phi_i}$ 
or, equivalently, the functions of
$\varepsilon$, $n_j$, and $w^{\mu}_{(i)}w_{(k) \mu}$.
%(see \cite{gusakov07} for more details).

In analogy with Ref.\ \cite{gusakov07},
from Eqs.\ (\ref{particle_conservation}), (\ref{energymomentum}), and (\ref{2ndlaw2})
one can derive
the entropy generation equation, which defines 
the dissipative corrections $\tau^{\mu \nu}$ and $\varkappa_i$, 
\begin{eqnarray}
\partial_{\mu} S^{\mu} &=&
- {\varkappa_i  \over T} \,\, \partial_{\mu} \left[ 
Y_{ik}
w^{\mu}_{(k)} \right] -\tau^{\mu \nu} \,\, \partial_{\mu} \left(
{u_{\nu} \over T} \right)
\nonumber \\
&&+ Y_{ik} w^{\mu}_{(k)} \,\, {\varkappa_i  \over T^2} \,\,
\partial_{\mu} T + u^{\nu} \,\, Y_{ik} w^{\mu}_{(k)} \,\,
{\varkappa_i \over T} \,\,
\partial_{\nu} u_{\mu}.  \qquad
\label{entropy2}
\end{eqnarray}
Here $S^{\mu}$ is the entropy current density,
\begin{equation}
S^{\mu} = S u^{\mu} 
- {u_{\nu} \over T} \,\, \tau^{\mu \nu} 
- {\varkappa_i \over T} \,\, Y_{ik} w^{\mu}_{(k)},
\label{S_mu}
\end{equation}
satisfying the natural constraint
$u_{\mu} S^{\mu} =-S$.
The last two terms in Eq.\ (\ref{entropy2}) 
are small and can be omitted in the majority of
applications (for more details, 
see the discussion in Ref.\ \cite{gusakov07}).

Using the requirement that the entropy does not decrease, 
one can obtain for the dissipative corrections 
(neglecting the last two terms 
in the right-hand side of Eq.\ \ref{entropy2})
\begin{eqnarray}
\tau^{\mu \nu} &=& - \kappa \, \left( H^{\mu \gamma} \, u^{\nu} +
H^{\nu \gamma} \, u^{\mu} \right) \left(  \partial_{\gamma} T + T
u^{\delta} \,
\partial_{\delta} u_{\gamma} \right)
\nonumber \\
&-& \eta \, H^{\mu \gamma} \, H^{\nu \delta} \,\, \left(
\partial_{\delta} u_{\gamma} + \partial_{\gamma} u_{\delta} - {2
\over 3} \,\, \eta_{\gamma \delta} \,\,
\partial_{\varepsilon} u^{\varepsilon}  \right)
\nonumber \\
&-& \xi_{1i} \, H^{\mu \nu} \,
\partial_{\gamma} \left[ Y_{ik} w^{\gamma}_{(k)} \right]
- \xi_{2} \, H^{\mu \nu} \, \partial_{\gamma} u^{\gamma},
\label{taumunu3}\\
\varkappa_{n} &=& - \xi_{3{i}} \, \partial_{\mu} \left[ Y_{ik}
w^{\mu}_{(k)} \right] - \xi_{4{n}} \, \partial_{\mu} u^{\mu},\\
\varkappa_{\rm \Lambda} &=& - \xi_{5{i}} \, \partial_{\mu} \left[
Y_{ik} w^{\mu}_{(k)} \right] - \xi_{4 {\rm \Lambda}} \,
\partial_{\mu} u^{\mu}, \\
\varkappa_{\rm \Sigma} &=& - \xi_{6{i}} \, \partial_{\mu} \left[
Y_{ik} w^{\mu}_{(k)} \right] - \xi_{4 {\rm \Sigma}} \,
\partial_{\mu} u^{\mu},\\
\varkappa_{p} &=& - \xi_{7{i}} \, \partial_{\mu} \left[ Y_{ik}
w^{\mu}_{(k)} \right] - \xi_{4 {p}} \,\partial_{\mu} u^{\mu}.
\label{diss}
\end{eqnarray}
Here $\kappa$ and $\eta$ are the thermal conductivity 
and shear viscosity coefficients, respectively;
$H^{\mu \nu} \equiv \eta^{\mu \nu}+u^\mu u^\nu$ 
is the projection matrix;
$\xi_{1i}, \xi_2, \xi_{3i}, \xi_{4i},
\xi_{5i}, \xi_{6i}$, and $\xi_{7i}$
are {\it twenty five} 
bulk viscosity coefficients ($i=n,p,\Lambda,\Sigma$).
We would like to emphasize that the dissipative corrections
(\ref{taumunu3})--(\ref{diss}) are incomplete in a sense that they 
($i$) do not include various terms
related to particle diffusion; 
($ii$) neglect (typically) small terms, 
explicitly depending on $w^{\mu}_{(i)}$. 
For example, we neglect the terms of the form 
$w^{\mu}_{(i)} \partial_{\mu} T$ and 
$u^{\mu} w^{\nu}_{(i)} \partial_{\gamma} u^{\gamma}$ 
in the expressions for
$\varkappa_{k}$ and $\tau^{\mu \nu}$, respectively.
The similar approximation is very well known in the non-relativistic theory
of superfluids (see, e.g., the textbook of Landau and Lifshitz \cite{ll87}, \S 140 
or Ref.\ \cite{gusakov07}).
An inclusion of all these dissipative terms would lead to a number of 
kinetic coefficients much larger than 27. 
Since in this paper we are mainly interested in the bulk viscosity coefficients, 
we restrict ourselves to a simplified form 
(\ref{taumunu3})--(\ref{diss}) of dissipative corrections.

The number of bulk viscosity coefficients can be reduced. 
Notice that, from the quasineutrality condition
(\ref{charge neutrality}) and 
the continuity equations (\ref{particle_conservation}), 
it follows that
\begin{equation}
\partial_{\mu} \left[ Y_{pk} w^{\mu}_{(k)}
\right] = \partial_{\mu} \left[ Y_{{\rm \Sigma}k} w^{\mu}_{(k)}
\right].
\label{neutral}
\end{equation}
A similar condition for superfluid {\it npe} matter was derived in 
Ref.\ \cite{gusakov07} (see Eq.\ 41 of that reference).
Owing to the condition (\ref{neutral}), 
there is no need to introduce
the bulk viscosity coefficients 
both for protons and for $\Sigma^-$ hyperons. 
For example, it is sufficient 
to introduce the quantity
$\xi_{1 \Sigma p} \equiv \xi_{1p} + \xi_{1\Sigma}$ 
instead of $\xi_{1p}$ and $\xi_{1\Sigma}$. 
Because of the same reason, 
we are not interested in the quantities $\varkappa_{\Sigma}$
and $\varkappa_{p}$ taken separately.
Instead, we will introduce the sum 
$\varkappa_{\Sigma p} \equiv 
\varkappa_{\Sigma}+\varkappa_{p}$ 
(notice that, $\varkappa_{\Sigma p}$ is the quantity that appears in 
the entropy generation equation \ref{entropy2}).
As a result, the corrections $\tau^{\mu \nu}$ and $\varkappa_q$ 
take the form
\begin{eqnarray}
\tau^{\mu \nu} &=& - \kappa \, \left( H^{\mu \gamma} \, u^{\nu} +
H^{\nu \gamma} \, u^{\mu} \right) \left(  \partial_{\gamma} T + T
u^{\delta} \,
\partial_{\delta} u_{\gamma} \right)
\nonumber \\
&-& \eta \, H^{\mu \gamma} \, H^{\nu \delta} \,\, \left(
\partial_{\delta} u_{\gamma} + \partial_{\gamma} u_{\delta} - {2
\over 3} \,\, \eta_{\gamma \delta} \,\,
\partial_{\varepsilon} u^{\varepsilon}  \right)
\nonumber \\
&-& \xi_{1q} \, H^{\mu \nu} \,
\partial_{\gamma} \left[ Y_{qk} w^{\gamma}_{(k)} \right]
- \xi_{2} \, H^{\mu \nu} \, \partial_{\gamma} u^{\gamma},
\label{tau_munu2} \\
\varkappa_{n} &=& - \xi_{3{q}} \, \partial_{\mu} \left[ Y_{qk}
w^{\mu}_{(k)} \right] - \xi_{4{n}} \, \partial_{\mu} u^{\mu},
\label{kappa_n} \\
\varkappa_{\Lambda} &=& - \xi_{5{q}} \, \partial_{\mu} \left[
Y_{qk} w^{\mu}_{(k)} \right] - \xi_{4 {\rm \Lambda}} \,
\partial_{\mu} u^{\mu}, \label{kappa_L} \\
\varkappa_{\Sigma p} &=& - \xi_{6{q}} \, \partial_{\mu} \left[
Y_{qk} w^{\mu}_{(k)} \right] - \xi_{4 {\Sigma p}} \,
\partial_{\mu} u^{\mu}.
\label{kappa_Sp}
\end{eqnarray}
Here and below the subscript $q$ takes on the values 
$n$, $\Lambda$, and $\Sigma p$.
%where the subscripts $i$ and $k$ 
%run over the particle species $n, \Lambda$, and $\Sigma p$. 
The expression $\partial_{\mu} \left[ Y_{\Sigma p k} w^{\mu}_{(k)}\right]$ 
in Eqs.\ (\ref{tau_munu2})--(\ref{kappa_Sp}) 
%one should understand 
means
$\partial_{\mu} \left[Y_{{\rm \Sigma}k} w^{\mu}_{(k)}\right]$ 
or, equivalently, $\partial_{\mu} \left[ Y_{pk} w^{\mu}_{(k)}\right]$.

As follows from the above equations, 
we have actually {\it sixteen} (rather than twenty five) 
bulk viscosity coefficients 
which can contribute to the dissipation of mechanical energy in
superfluid nucleon-hyperon matter.
Using the Onsager symmetry principle, we obtain
%only ten of them are independent
%
\begin{eqnarray}
\xi_{3\Lambda}=\xi_{5n},\; \xi_{3\Sigma p}=\xi_{6n},\;
\xi_{4n}=\xi_{1n},\; \xi_{5\Sigma p}=\xi_{6\Lambda},\;
\xi_{4\Lambda}=\xi_{1\Lambda},\; \xi_{4\Sigma p}=\xi_{1\Sigma p}.
\label{onsager11}
\end{eqnarray}
%
%only ten of them are independent.
Thus, generally, only ten of them are independent.

%%%%%%%%%%%%%%%%%%%%%%%%%%%%%%%%%%%%%%%%%%%%%%%%%%%%%%%
\subsection{Calculation of the bulk viscosity coefficients
for superfluid nucleon-hyperon matter}
%%%%%%%%%%%%%%%%%%%%%%%%%%%%%%%%%%%%%%%%%%%%%%%%%%%%%%%

Let us calculate these phenomenological coefficients 
assuming they are due to non-equilibrium 
processes (\ref{s})--(\ref{ll}).
As in Sec.\ II, we assume that the matter is slightly perturbed
out of equilibrium state and pulsates with the frequency $\omega$.
Since the deviation from the equilibrium is small the hydrodynamic equations 
can be linearized. Because of the same reason, the dependence of 
various thermodynamic quantities (e.g., the pressure $P$) 
on the scalars $w^{\mu}_{(i)} w_{(k) \mu}$ can be neglected. 
We assume that the four-vectors $w^{\mu}_{(i)}$ 
characterizing the superfluid flow of particle species 
$i=n,p,\Lambda$, or $\Sigma$
equal zero in the unperturbed matter. 
Thus, by perturbing the system, we produce some small $w^{\mu}_{(i)}$ 
so that the scalars $w^{\mu}_{(i)} w_{(k) \mu}$ 
will be of the second-order smallness 
and can be omitted in the linear approximation.
%in superfluid nucleon-hyperon matter.

We start from the non-dissipative 
relativistic hydrodynamics of superfluid nucleon-hyperon mixture.
%assuming that there are no dissipative processes, 
%except for the non-equilibrium reactions.
In this case 
the energy-momentum tensor 
is given by Eq.\ (\ref{Tmunu2}), 
where the dissipative correction $\tau^{\mu\nu}$ 
should be set zero.
Similarly, the expressions for $w_{(i)}^{\mu}$
are given by Eq.\ (\ref{w_i2}) with $\varkappa_i=0$.
The non-equilibrium processes (\ref{s})--(\ref{ll}) lead 
to the appearance of the sources in the right-hand side of the continuity 
equations (\ref{particle_conservation}) 
which, as we will demonstrate below, 
generate the `effective' dissipative corrections $\tau^{\mu\nu}$ and 
$\varkappa_q$.

To calculate $\xi_{1q}$ and $\xi_2$
%generated by the non-equilibrium reactions (\ref{s})--(\ref{ll}), 
it is convenient to expand the energy-momentum tensor 
of the pulsating matter (\ref{Tmunu2}) 
(with $\tau^{\mu\nu} =  0$) in the comoving frame 
(where $u^\mu=(1,0,0,0)$)
near the equilibrium, 
as it was done in Ref.\ \cite{gusakov07},
\begin{eqnarray}
&& T^{00} = \varepsilon_0 + \delta \varepsilon,
\nonumber \\
&& T^{0m}= T^{m0} =
\mu_{i0} Y_{ik} \, w^{m}_{(k)},
\nonumber \\
&& T^{nm} = \left( P_0
+ \delta P \right) \,\, \delta_{nm}.
\label{Tmunu3}
\end{eqnarray}
Here we restrict ourselves to the linear perturbation terms.
The spatial indices $n$ and $m$ vary over $1,2,$ and $3$; 
$\varepsilon_0, \mu_{i0}$, and $P_0$ 
are the corresponding thermodynamic quantities
calculated at equilibrium
(in the absence of pulsations).

Now our aim is to extract from the 
tensor (\ref{Tmunu3}) various terms which are generated
by the non-equilibrium processes (\ref{s})--(\ref{ll}) and contribute
to dissipation. 
Then, we will write these terms
in the form of a separate dissipative tensor 
$\tau^{\mu \nu}_{\rm bulk}$ .
%Our consideration essentially follows
%that of Ref.\ \cite{gusakov07}. 

As follows from Eq.\ (\ref{2ndlaw3}), in the linear approximation 
$\delta \varepsilon$ 
remains the same as in the absence of dissipation,
$\delta \varepsilon = \mu_n \delta n_b$
(this is because the dissipative processes (\ref{s})--(\ref{ll}) 
conserve the number of baryons, 
hence $\delta n_b$ is independent 
of the reaction rates $\lambda_l$, $l=$1,2,3, or 4).
Thus, the component $\tau^{00}_{\rm bulk}$ 
of the tensor $\tau^{\mu \nu}_{\rm bulk}$ is zero. 
Similarly, $\tau^{m 0}_{\rm bulk}=\tau^{0 m}_{\rm bulk}=0$.
On the contrary,
the variation $\delta P$ of the pressure contains
a dissipative part (we denote it by $\delta P_{\rm dis}$). 
According to Refs.\ \cite{hly00,lo02} and Sec.\ 2, it is given by
\begin{equation}
\delta P_{\rm dis} = {\rm Re}(\delta P),
\label{dPdis}
\end{equation}
%
%Denoting this part of $\delta P$ as $\delta P_{\rm dis}$ 
so that 
the tensor $\tau^{\mu \nu}_{\rm bulk}$ 
can be presented in the form
(in the comoving frame)
%
%Keeping only the terms contributing to dissipation 
%(see Ref.\ \cite{gusakov07}), 
%one obtains the dissipative correction to 
%the energy-momentum tensor
%
\begin{eqnarray}
&& \tau^{00}_{\rm bulk} =0,
\nonumber \\
&& \tau^{0m}_{\rm bulk}=\tau^{m0}_{\rm bulk}=0,
\nonumber \\
&& \tau^{nm}_{\rm bulk}=\delta P_{\rm dis} \,\, \delta_{nm}.
\label{tau_bulk}
\end{eqnarray}
%
%where $\delta P_{\rm dis}$ is the part of the pressure variation $\delta P$
%which is ({\it i}) due to the non-equilibrium reactions
%(\ref{s})-(\ref{ll}) and, in addition, 
%({\it ii}) responsible for the dissipation.
%
Let us determine $\delta P_{\rm dis}$.
%
%For that, 
%we will expand the pressure $P$ 
%near the equilibrium state as a function of $n_b$, $n_H$, 
%$n_{\Sigma n} \equiv n_{\Sigma} + n_n$,  and $y \equiv x_e/x_{\mu}$.
%
For that purpose, we present the pressure $P$ as a function of 
$n_b$, $n_H$, 
$n_{\Sigma n} \equiv n_{\Sigma} + n_n$,  and $y \equiv x_e/x_{\mu}$.
All other number densities can be expressed through 
$n_b$, $n_H$, $n_{\Sigma n}$, and $y$ with the help of 
Eqs.\ (\ref{eqfast}) and (\ref{charge neutrality}).
Notice that we choose $n_{\Sigma n}$ and $y$
instead of the variables $x_e$ and $x_{\mu}$ of Sec.\ 2. 
The variables $x_e$ and $x_{\mu}$ are less convenient here
because Eq.\ (\ref{dxedxm}) does not hold 
in the case of superfluid matter.
Expanding the pressure $P(n_b, n_H, n_{\Sigma n}, y)$ 
near the equilibrium state, 
%as a function of $n_b$, $n_H$, $n_{\Sigma n}$, and $y$, 
we write
\begin{equation}
\delta P_{\rm dis} =  \frac{\partial P}{\partial n_{b}} \, {\rm Re} (\delta n_b)
+ \frac{\partial P}{\partial n_{H}} \, {\rm Re} (\delta n_H)
+\frac{\partial P}{\partial n_{\Sigma n}} \, {\rm Re} (\delta n_{\Sigma n})
+\frac{\partial P}{\partial y} \, {\rm Re} (\delta y).
\label{dPdiss2}
\end{equation}
%

%All other number densities can be expressed through 
%$n_b$, $n_H$, $n_{\Sigma n}$, and $y$ with the help of 
%Eqs.\ (\ref{eqfast}) and (\ref{charge neutrality}).
It is straightforward to show that 
\begin{equation}
\delta y =0,
\label{dy}
\end{equation}
as a consequence of the continuity equations 
(\ref{ne}) and (\ref{nmu}) for electrons and muons.
The variations of other variables, $n_b$, $n_H$, and $n_{\Sigma n}$,
can be found from corresponding continuity equations.
Using Eq.\ (\ref{particle_conservation}) and the fact,
that 
%the number density 
the variations 
depend on time $t$ as ${\rm exp}(i \omega t)$, 
we obtain in the comoving frame
\begin{eqnarray}
%\frac{\partial\delta n_{b}}{\partial t}
i \omega \, \delta n_{b}
+{\rm div}({\pmb J}_b)=0, 
\label{b}\\
%\frac{\partial\delta n_{H}}{\partial t}
i \omega \, \delta n_{H}
+{\rm div}
({\pmb J}_H)=\Delta \Gamma,
\label{H} \\
%\frac{\partial\delta n_{\Sigma n}}{\partial t}
i \omega \, \delta n_{\Sigma n}
+{\rm div}
({\pmb J}_{\Sigma n})=-\Delta \Gamma.
\label{Sn}
\end{eqnarray}
Here
$\Delta \Gamma \equiv \Delta \Gamma_1 + 
\Delta \Gamma_2+\Delta \Gamma_3+\Delta \Gamma_4 = \lambda \, \delta \mu$,
and 
\begin{equation}
\delta \mu(n_b,n_H,n_{\Sigma n}) = 
\frac{\partial \delta \mu}{\partial n_{b}} \, \delta n_b
+\frac{\partial \delta \mu}{\partial n_{H}} \, \delta n_H
+\frac{\partial \delta \mu}{\partial n_{\Sigma n}} \, \delta n_{\Sigma n}.
\label{dmu2}
\end{equation}
In Eqs.\ (\ref{b})--(\ref{Sn})
$\pmb{J}_{b}\equiv n_{b} \pmb{u}+\sum_i 
Y_{ik}\pmb{w}_{(k)}$; $\pmb{J}_{H}\equiv n_{H} \pmb{u}
+Y_{\Sigma k}\pmb{w}_{(k)}
+Y_{\Lambda k}\pmb{w}_{(k)}$;
$\pmb{J}_{\Sigma n}\equiv n_{{\Sigma n}} 
\pmb{u} +Y_{\Sigma k}\pmb{w}_{(k)}+Y_{nk}
\pmb{w}_{(k)}$; 
${\pmb u}$ and ${\pmb w}_{(i)}$ are 
the spatial components of the four-velocity $u^{\mu}$ 
and four-vector $w^{\mu}_{(i)}$, respectively.
The solution to the above system of equations gives
\begin{eqnarray}
{\rm Re}(\delta n_{b}) &=&0,
\label{dnHSn1} \\
{\rm Re}(\delta n_{\Sigma n}) &=& -{\rm Re}(\delta n_{H}),
\label{dnHSn2} \\
{\rm Re}(\delta n_{H})&=&
%-{\rm Re}(\delta n_{\Sigma n})=
%\frac{1}{\lambda}\left(\frac{\partial \delta \mu}{\partial 
%n_{H}}-\frac{\partial \delta \mu}{\partial 
%n_{\Sigma n}}\right)^{-2}\frac{1}{\omega^2\tau^2+1} 
%\nonumber \\
%&& \times
k \, 
\left[ \frac{\partial \delta \mu}{\partial n_{b}}{\rm div} 
(\pmb{J}_{b})+ \frac{\partial \delta \mu}{\partial 
n_{H}} {\rm div} (\pmb{J}_{H})
+ \frac{\partial \delta \mu}{\partial
n_{\Sigma n}}{\rm div} (\pmb{J}_{\Sigma n}) \right],
\label{dnHSn}
\end{eqnarray}
where we use the notations $1/k \equiv \lambda\left(\partial \delta
\mu/\partial n_{H}-\partial \delta \mu/\partial 
n_{\Sigma n}\right)^{2}(1+\omega^2\tau^2)$; 
$\tau \equiv 1/\lambda  
\;(\partial \delta \mu/\partial n_{H}
-\partial \delta \mu /\partial n_{\Sigma n} )^{-1}$.
Now, using Eq.\ (\ref{dPdiss2}) for $\delta P_{\rm dis}$
and Eqs.\ (\ref{dnHSn1})--(\ref{dnHSn}), 
we can find the dissipative tensor 
$\tau^{\mu \nu}_{\rm bulk}$ in the comoving frame (see Eq.\ (\ref{tau_bulk})).
Then it can be easily rewritten in an arbitrary frame. 
The result is
\begin{eqnarray}
\tau^{\mu \nu}_{\rm bulk} &=& H^{\mu \nu} \, k \, 
\left( 
\frac{\partial P}{\partial n_H} - \frac{\partial P}{\partial n_{\Sigma n}}
\right) \, 
\left[
\left(n_b \, \frac{\partial \delta \mu}{\partial n_b}
+ n_H \, \frac{\partial \delta \mu}{\partial n_H}
+n_{\Sigma n} \, \frac{\partial \delta \mu}{\partial n_{\Sigma n}}
\right) \, \partial_{\gamma} u^{\gamma}
\right.
\nonumber \\
&& \left. + \left( \frac{\partial \delta \mu}{\partial n_b} 
+\frac{\partial \delta \mu}{\partial n_{\Sigma n}} \right) \, 
\partial_{\gamma} \left( Y_{nk}w^{\gamma}_{(k)} \right)
+ \left(2 \frac{\partial \delta \mu}{\partial n_{b}} 
+ \frac{\partial \delta \mu}{\partial n_{H}} 
+ \frac{\partial \delta \mu}{\partial n_{\Sigma n}}\right) \, 
\partial_{\gamma} \left( Y_{\Sigma p k}w^{\gamma}_{(k)} \right)
\right.
\nonumber \\
&& \left. 
+ \left(  \frac{\partial \delta \mu}{\partial n_{b}} 
+ \frac{\partial \delta \mu}{\partial n_{H}} \right) \,
\partial_{\gamma} \left( Y_{\Lambda k}w^{\gamma}_{(k)} \right)
\right].
\label{taumunu}
\end{eqnarray}
A comparison of $\tau^{\mu \nu}_{\rm bulk}$ with
the phenomenological dissipative tensor $\tau^{\mu \nu}$ 
(see Eq.\ (\ref{tau_munu2}))
allows us to identify the bulk viscosity coefficients
$\xi_{1n}$, $\xi_{1 \Lambda}$, $\xi_{1 \Sigma p}$, and $\xi_2$,
generated by non-equilibrium processes (\ref{s})--(\ref{ll})

\begin{eqnarray}
\xi_{1 n} &=& -k \, \left(\frac{\partial P}
{\partial n_{H}}-\frac{\partial P}{\partial 
n_{\Sigma n}}\right)  
\left(\frac{\partial \delta \mu}{\partial n_{b}}+\frac{\partial \delta
\mu}{\partial n_{
\Sigma n}}\right), 
\label{xi1n} \\
\xi_{1 \Lambda} &=& -k \, \left(\frac{\partial P}
{\partial n_{H}}-\frac{\partial P}{\partial 
n_{\Sigma n}}\right)
\left(\frac{\partial \delta \mu}{\partial n_{b}}+\frac{\partial
\delta \mu}{\partial n_{H}}\right),
\label{xi1L} \\
\xi_{1 \Sigma p} &=& -k \, \left(\frac{\partial P}
{\partial n_{H}}-\frac{\partial P}{\partial 
n_{\Sigma n}}\right) 
\left(2 \frac{\partial \delta \mu}{\partial n_{b}}+\frac{\partial
\delta \mu}{\partial n_{H}}
+ \frac{\partial \delta \mu}{\partial n_{\Sigma n}}\right),
\label{xiSp} \\
\xi_2 &=& -k \left(\frac{\partial P}
{\partial n_{H}}-\frac{\partial P}{\partial 
n_{\Sigma n}}\right)\left(n_{b}\frac{\partial \delta \mu}{\partial 
n_{b}}+n_{H}\frac{\partial \delta \mu}{\partial n_{H}}
+n_{\Sigma n}\frac{\partial \delta \mu}{\partial 
n_{\Sigma n}}\right).
\label{xi22}
\end{eqnarray}
It can be shown, that the expression 
for $\xi_2$ formally coincides with Eq.\ (\ref{ksi}) 
for the bulk viscosity of normal matter 
(however, these formulae give 
different numerical results because the reaction rates 
$\lambda_l$ ($l=1$,$\ldots$,$4$) differ for superfluid and normal matter).  
To prove this, we can rewrite Eq.\ (\ref{xi22}) using 
$n_{b}$, $x_{H}$, $x_e$, and $x_{\mu}$
as independent variables 
instead of $n_b$, $n_H$, $n_{\Sigma n}$, and $y$.
%When doing this it is useful 
%to note that for any given function $f(n_b, x_H, x_e, x_\mu)$
%the following equalities must be satisfied
The following equalities will be helpful 
($f$ is an arbitrary function)
\begin{eqnarray}
\frac{1}{n_b} \, \frac{\partial f(n_{b}, x_{H}, x_e, x_{\mu})}{\partial x_{H}} &=&
\frac{\partial f(n_{b},n_{H},n_{\Sigma n}, y)}{\partial
n_{H}}-\frac{\partial f(n_{b},n_{H},n_{\Sigma n}, y)}
{\partial n_{\Sigma n}}, \\
n_{b}\frac{\partial f(n_{b},
x_{H}, x_e,x_{\mu})}{\partial n_{b}} &=&
n_{b}\frac{\partial f(n_{b},n_{H},n_{\Sigma
n},y)}{\partial n_{b}}
\nonumber\\
&&+n_{H}\frac{\partial f(n_{b},
n_{H},n_{\Sigma n},y)}{\partial n_{H}} +n_{\Sigma
n}\frac{\partial f(n_{b},n_{H},n_{\Sigma n},y)}{\partial
n_{\Sigma n}}.
\end{eqnarray}

\begin{figure}
\begin{center}
\leavevmode \epsfxsize=8cm \epsfbox[60 200 560 670]{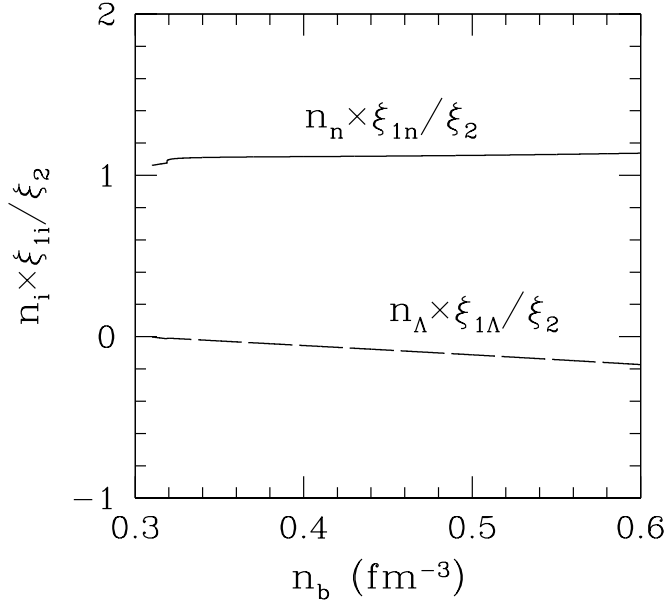}
\end{center}
\caption{ Dimensionless parameters
$\xi_{1i} n_i/\xi_2$ ($i=n, \Lambda$) 
%normalized to $\xi_2/n_i$ 
versus 
%baryon number density
$n_{b}$ in superfluid nucleon-hyperon matter.
} 
\label{fig:fig2}
\end{figure}

To calculate other bulk viscosity coefficients let us 
apply the same consideration to Eq.\ (\ref{w_i2}) for $w^{\mu}_{(i)}$. 
As a result, we obtain (in the comoving frame) the
expression for the dissipative component $\varkappa_i$ generated by 
nonequilibrium processes (\ref{s})--(\ref{ll})
\begin{equation}
\varkappa_i=\delta {\mu_i}_{\rm dis}=
\frac{\partial \mu_i}{\partial n_{b}} 
{\rm Re}(\delta n_{b})+
\frac{\partial \mu_i}{\partial n_{H}} 
{\rm Re}(\delta n_{H})+\frac{\partial \mu_i}{\partial n_{\Sigma n}} 
{\rm Re}(\delta n_{\Sigma n}). 
\label{kappa_i}
\end{equation}
Here $\delta \mu_{i {\rm dis}}$ is the dissipative term in the Taylor expansion of
the chemical potential $\mu_i$ near the equilibrium state 
(similar to $\delta P_{\rm dis}$); 
${\rm Re}(\delta n_{b})$,
${\rm Re}(\delta n_{H})$, and ${\rm Re}(\delta n_{\Sigma n})$ 
are taken from Eqs.\ (\ref{dnHSn1})--(\ref{dnHSn}).
In a fully covariant form, $\varkappa_q$ is given by 
Eqs.\ (\ref{kappa_n})--(\ref{kappa_Sp}) 
with the bulk viscosity coefficients
\begin{eqnarray}
\xi_{3 n} &=& -k \,
\left(     
\frac{\partial \mu_{n}}{\partial n_{H}}  
-\frac{\partial \mu_{n}}{\partial n_{\Sigma n}}    
\right)
\left( \frac{\partial \delta \mu}{\partial n_{b}}
+\frac{\partial \delta \mu}{\partial n_{\Sigma n}} \right), 
\label{beg}\\
\xi_{3 \Lambda} &=& -k \, 
\left(     
\frac{\partial \mu_{n}}{\partial n_{H}}  
-\frac{\partial \mu_{n}}{\partial n_{\Sigma n}}    
\right)
\left(\frac{\partial \delta
\mu}{\partial n_{b}}
+\frac{\partial \delta \mu}{\partial n_{H}}\right), 
\\
\xi_{3 \Sigma p} &=& -k \,
\left(     
\frac{\partial \mu_{n}}{\partial n_{H}}  
-\frac{\partial \mu_{n}}{\partial n_{\Sigma n}}    
\right)
\left(
2 \frac{\partial \delta \mu}{\partial n_{b}}
+\frac{\partial \delta \mu}{\partial n_{H}} 
+ \frac{\partial \delta \mu}{\partial n_{\Sigma n}}
\right), 
\\
\xi_{4 n} &=& -k \, 
\left(     
\frac{\partial \mu_{n}}{\partial n_{H}}  
-\frac{\partial \mu_{n}}{\partial n_{\Sigma n}}    
\right)
\left(
n_{b} \frac{\partial \delta \mu}{\partial n_{b}}
+n_{H}\frac{\partial \delta \mu}{\partial n_{H}} 
+n_{\Sigma n}\frac{\partial \delta \mu}{\partial n_{\Sigma n}}
\right),
\\
\xi_{4 \Lambda} &=& -k \, 
\left(     
\frac{\partial \mu_{\Lambda}}{\partial n_{H}}  
-\frac{\partial \mu_{\Lambda}}{\partial n_{\Sigma n}}    
\right)
\left(
n_{b} \frac{\partial \delta \mu}{\partial n_{b}}
+n_{H}\frac{\partial \delta \mu}{\partial n_{H}} 
+n_{\Sigma n}\frac{\partial \delta \mu}{\partial n_{\Sigma n}}
\right), 
\\
\xi_{4 \Sigma p} &=& -k \, 
\left[    
\frac{\partial (\mu_{p}+ \mu_{\Sigma})}{\partial n_{H}}  
-\frac{\partial (\mu_{p}+\mu_{\Sigma})}{\partial n_{\Sigma n}}    
\right]
\left(
n_{b} \frac{\partial \delta \mu}{\partial n_{b}}
+n_{H}\frac{\partial \delta \mu}{\partial n_{H}} 
+n_{\Sigma n}\frac{\partial \delta \mu}{\partial n_{\Sigma n}}
\right), 
\\
\xi_{5 n} &=& -k \, 
\left(     
\frac{\partial \mu_{\Lambda}}{\partial n_{H}}  
-\frac{\partial \mu_{\Lambda}}{\partial n_{\Sigma n}}    
\right)
\left(\frac{\partial \delta \mu}{\partial n_{b}}
+\frac{\partial \delta \mu}{\partial n_{\Sigma n}}\right), 
\\
\xi_{5 \Lambda} &=& -k \, 
\left(     
\frac{\partial \mu_{\Lambda}}{\partial n_{H}}  
-\frac{\partial \mu_{\Lambda}}{\partial n_{\Sigma n}}    
\right)
\left(\frac{\partial \delta \mu}{\partial n_{b}}
+\frac{\partial \delta \mu}{\partial n_{H}}\right), 
\\
\xi_{5 \Sigma p} &=& -k \, 
\left(     
\frac{\partial \mu_{\Lambda}}{\partial n_{H}}  
-\frac{\partial \mu_{\Lambda}}{\partial n_{\Sigma n}}    
\right)
\left(2 \frac{\partial \delta \mu}{\partial n_{b}} 
+\frac{\partial \delta \mu}{\partial n_{H}}+
\frac{\partial \delta \mu}{\partial n_{\Sigma n}}\right), 
\\
\xi_{6 n} &=& -k \, 
\left[    
\frac{\partial (\mu_{p}+ \mu_{\Sigma})}{\partial n_{H}}  
-\frac{\partial (\mu_{p}+\mu_{\Sigma})}{\partial n_{\Sigma n}}    
\right]
\left(\frac{\partial \delta \mu}{\partial n_{b}}
+\frac{\partial \delta \mu}{\partial n_{\Sigma n}}\right), 
\\
\xi_{6 \Lambda} &=& -k \, 
\left[    
\frac{\partial (\mu_{p}+ \mu_{\Sigma})}{\partial n_{H}}  
-\frac{\partial (\mu_{p}+\mu_{\Sigma})}{\partial n_{\Sigma n}}    
\right]
\left(\frac{\partial \delta \mu}{\partial n_{b}}
+\frac{\partial \delta \mu}{\partial n_{H}}\right), 
\\
\xi_{6 \Sigma p} &=& -k \, 
\left[    
\frac{\partial (\mu_{p}+ \mu_{\Sigma})}{\partial n_{H}}  
-\frac{\partial (\mu_{p}+\mu_{\Sigma})}{\partial n_{\Sigma n}}    
\right]
\left(
2 \frac{\partial \delta \mu}{\partial n_{b}}  
+\frac{\partial \delta \mu}{\partial n_{H}}
+\frac{\partial \delta \mu}{\partial n_{\Sigma n}}
\right).
\label{fin}
\end{eqnarray}
In Eqs.\ (\ref{beg})--(\ref{fin}) we assumed that 
the thermodynamic quantities
are functions of $n_b$, $n_H$, $n_{\Sigma n}$, and $y$.
One can easily check that 
all the six Onsager relations (\ref{onsager11}) are satisfied.
Moreover, it turns out that the bulk viscosities 
(\ref{xi1n})--(\ref{xi22}) and (\ref{beg})--(\ref{fin})
obey a number of additional constraints ($q=n$, $\Lambda$, $\Sigma p$)
\begin{eqnarray}
\xi_{6 q}&=&\xi_{3q}+\xi_{5q},\qquad
\xi_{4 \Sigma p}= \xi_{4 n}+\xi_{4 \Lambda},
\label{Sp} \\
\xi_{1 n}^2 &=& \xi_2 \xi_{3 n},\; \xi_{1 \Lambda}^2=\xi_2
\xi_{5 \Lambda},\; \xi_{1 \Sigma p}^2=\xi_2 \xi_{6 \Sigma p}, 
\label{additional1} \\
\xi_{1 n}\xi_{1 \Lambda}&=& \xi_2 \xi_{5 n},\; \xi_{1
\Sigma p}\xi_{1 \Lambda}=\xi_2 \xi_{6 \Lambda},\; \xi_{1
n}\xi_{1 \Sigma p}=\xi_2 \xi_{6 n}
\label{additional2}
\end{eqnarray} 
so that we have only {\it three} independent bulk viscosity coefficients, 
say $\xi_2, 
\xi_{1n}$, and $\xi_{1 \Lambda}$.
All other coefficients can be expressed through 
these three.
The coefficients $\xi_2, \xi_{1n}$, 
and $\xi_{1 \Lambda}$ are compared in Fig.\ \ref{fig:fig2}.
Because the dimensions of the 
coefficients $\xi_{1i}$ ($i=n, \Lambda$) and $\xi_2$ are different,
we show the dimensionless combinations 
$\xi_{1i} n_i/\xi_2$ 
%($i=n, \Lambda$) 
as functions of the baryon number density $n_{b}$.

%%%%%%%%%%%%%%%%%%%%%%%%%%%%%%%%%%%%%%%%%%%%%%%%%%%%%%%%%%%%%%%%%%%%
%%%%%%%%%%%%%%%%%%%%%%%%%%%%%%%%%%%%%%%%%%%%%%%%%%%%%%%%%%%%%%%%%%%%
%
What is the nature of the relations (\ref{Sp})--(\ref{additional2})?
The relations (\ref{Sp}) follow from the equilibrium condition (\ref{eqfast}) 
with respect to the fast reaction (\ref{fast}).
This condition is valid even if we allow for the dissipative processes 
(\ref{s})--(\ref{ll}). 
Consequently, we can write
\begin{equation}
{\rm Re}(\mu_{\Sigma})+{\rm Re}(\mu_p) = {\rm Re}(\mu_n)+{\rm Re}(\mu_{\Lambda})
\label{Remu}
\end{equation}
or, in view of Eq.\ (\ref{kappa_i}) 
(we remind that ${\rm Re}(\mu_i) \equiv \mu_{i {\rm dis}}$; 
$\varkappa_{\Sigma p} \equiv \varkappa_{\Sigma}+\varkappa_p$),
\begin{equation}
\varkappa_{\Sigma p} = \varkappa_{n}+\varkappa_{\Lambda}.
\label{kappa}
\end{equation}
Then, substituting Eqs.\ (\ref{kappa_n})--(\ref{kappa_Sp}) into Eq.\ (\ref{kappa}) 
and equating prefactors in front of the same four-divergences, 
we obtain the relations (\ref{Sp}).

It is convenient to note that 
the constraint (\ref{kappa}) leads to a simple equation 
interrelating four-vectors $w^{\mu}_{(i)}$.
It holds true only if the bulk viscosities 
are generated by the non-equilibrium reactions.
%
%To derive this equation we rewrite
%the potentiality condition (\ref{w_i}) 
%in the following equivalent form 
%(compare it with the similar equation 31 of Ref.\ \cite{gusakov07})
%
%\begin{eqnarray}
%\partial^{\nu} \left[ w^{\mu}_{(i)}
%+q_i A^{\mu} + (\mu_i + \varkappa_i) u^{\mu} \right]
%= \partial^{\mu} \left[ w^{\nu}_{(i)}
% +q_i A^{\nu} +(\mu_i + \varkappa_i) u^{\nu} \right].
%\label{wmu22}
%\end{eqnarray}
%
%Summing up these conditions (with a proper sign) 
%for all baryon species and using  
%Eqs.\ (\ref{eqfast}) and (\ref{kappa}), one gets
To derive this equation, we sum up the potentiality conditions (\ref{w_i2})
(with proper signs) for all baryon species. 
Then, using  Eqs.\ (\ref{eqfast}) and (\ref{kappa}), we get 
\begin{eqnarray}
\partial^{\nu} \left[ w^{\mu}_{(n)}+w^{\mu}_{(\Lambda)}-w^{\mu}_{( 
\Sigma)}-w^{\mu}_{(p)}
\right]
= \partial^{\mu} \left[ w^{\nu}_{(n)}+w^{\nu}_{(\Lambda)}-w^{\nu}_{( 
\Sigma)}-w^{\nu}_{(p)} \right].
\end{eqnarray}
In view of this equation and Eq.\ (\ref{neutral}), there are only $4-2=2$
independent four-vectors $w^{\mu}_{(i)}$ in the system.

Now let us explain why the bulk viscosities 
satisfy Eqs.\ (\ref{additional1}) and
(\ref{additional2}).
For that purpose, we consider the entropy generation equation.
Neglecting all the dissipative processes except for the 
non-equilibrium reactions (\ref{s})--(\ref{ll}) 
(e.g., neglecting the thermal conductivity and shear viscosity),
we can obtain from the hydrodynamics discussed in this section,
\begin{equation}
T \, \partial_{\mu} S^{\mu} = 
\delta \mu \, \Delta \Gamma = \lambda \, \delta \mu^2.
\label{entropy3}
\end{equation}
A similar expression is valid for ${\it npe}$ matter 
(see Eq.\ (79) of Ref.\ \cite{gusakov07}).
The chemical potential disbalance $\delta \mu$ is given by 
Eq.\ (\ref{dmu2}). 
Substituting into Eq.\ (\ref{dmu2}) 
variations $\delta n_b$, $\delta n_{H}$, 
and $\delta n_{\Sigma n}$ calculated from
the continuity equations (\ref{b})--(\ref{Sn}), we find
\begin{equation}
\delta \mu = 
%\frac{\partial\delta\mu}{\partial n_{b}} \delta n_{b} + 
%\frac{\partial\delta\mu}{\partial n_{H}} \delta n_{H} + 
%\frac{\partial\delta\mu}{\partial n_{\Sigma n}} \delta n_{\Sigma n} 
%\nonumber \\
\frac{1}{\lambda \; \left(\partial\delta\mu/\partial n_{H} -
\partial\delta\mu/\partial n_{\Sigma n} \right)
- i \omega}\left[\frac{\partial\delta\mu}{\partial n_{b}} {\rm div} 
(\pmb{J}_{b}) + \frac{\partial\delta\mu}{\partial n_{H}} {\rm div} 
(\pmb{J}_{H}) + \frac{\partial\delta\mu}{\partial n_{\Sigma n}} {\rm 
div} (\pmb{J}_{\Sigma n}) \right].
\label{dmu}
\end{equation}
It follows from Eq.\ (\ref{dmu}), that for any given $u^{\mu}$ 
it is always possible to choose four-vectors $w^{\mu}_{(i)}$
in such a way that $\delta\mu=0$ 
at some point and in some particular moment 
(even if some baryon species are non-superfluid).
In other words, this means that we can vanish 
the entropy generation rate (\ref{entropy3}) at this point.

On the other hand, the entropy generation equation in terms of 
the effective bulk viscosities 
takes the form (see Eqs.\ (\ref{entropy2}) and (\ref{taumunu3})--(\ref{diss}))
\begin{eqnarray}
T \;\partial_{\mu} S^{\mu} =\left\{ \xi_{1q} \partial_\mu [Y_{qk} w^\mu_{(k)}] + 
\xi_2 \partial_\mu u^\mu\right\}\partial_\mu u^\mu+ \nonumber \\
\left\{ \xi_{3q} \partial_\mu [Y_{qk} w^\mu_{(k)}] + \xi_{4n} \partial_\mu 
u^\mu \right\}\partial_\mu [Y_{n k} w^\mu_{(k)}]+ \nonumber \\
\left\{ \xi_{5q} \partial_\mu [Y_{qk} w^\mu_{(k)}] + \xi_{4\Lambda} 
\partial_\mu u^\mu \right\}\partial_\mu [Y_{{\Lambda} k} w^\mu_{(k)}]+ 
\nonumber \\
\left\{ \xi_{6q} \partial_\mu [Y_{qk} w^\mu_{(k)}] + \xi_{4 \Sigma p} 
\partial_\mu u^\mu\right\}\partial_\mu [Y_{\Sigma p k} w^\mu_{(k)}].
\label{entropy4}
\end{eqnarray}
The right-hand side 
of this equation satisfies two conditions.
First, it is a positive-definite quadratic form 
(entropy cannot decrease!).
Second, according to Eqs.\ (\ref{entropy3}) and (\ref{dmu}), 
one can vanish it by an appropriate choice of 
$u^{\mu}$ and $w^{\mu}_{(i)}$
(at some particular moment and at some point).
There is a mathematical theorem that these two conditions 
are consistent with each other if and only if the determinant 
of the matrix composed of coefficients of the quadratic form vanishes.
This result is independent of an actual number of superfluid baryon species.
That is, the determinant will be zero in the case when 
all four baryon species are superfluid
as well as in the case when some of them are normal 
(for example, nucleons are normal, hyperons are superfluid, 
or neutrons are superfluid, while other particles are normal). 
%altogether, 7 combinations).                           
As a consequence, we arrive at the six additional constraints 
(\ref{additional1}) and (\ref{additional2})
on the bulk viscosity coefficients.

In this section we have assumed that baryons of all species are superfluid.
However, the hydrodynamics formulated here can be easily extended
to the situation when some baryon species are normal. 
In this case, one should vanish
matrix elements $Y_{ik}$ related to these baryon species.
For example, if neutrons are non-superfluid, then $Y_{nk} =Y_{kn} \equiv 0$.

%%%%%%%%%%%%%%%%%%%%%%%%%%%%%%%%%%%%%%%%%%%%%%%%%%%%%%%%%%%%%%%%%%%%%%%%%%%%%%%%
\section{Summary}
%%%%%%%%%%%%%%%%%%%%%%%%%%%%%%%%%%%%%%%%%%%%%%%%%%%%%%%%%%%%%%%%%%%%%%%%%%%%%%%%

In this paper we have analyzed the bulk viscosity 
due to non-equilibrium 
particle transformations in superfluid 
nucleon-hyperon matter of neutron stars.
Our approach is similar to that 
used in Ref.\ \cite{gusakov07} 
for the case of superfluid {\it npe} matter.

Our main results are as follows:

({\it i}) We have demonstrated that the expression for the bulk 
viscosity of normal (non-superfluid) nucleon-hyperon matter,
widely used in the literature, is inaccurate.
We have presented the correct derivation of the bulk viscosity 
and compared it with the results of Ref.\ \cite{lo02}.
Numerically, both formulae give almost similar results 
(within a few tens of percent).

({\it ii}) We have extended the hydrodynamics of superfluid mixtures
reported in Refs.\ \cite{ga06,gusakov07} 
to allow for a possible presence of superfluid hyperons.
We have determined the general form of dissipative terms 
entering the equations of this hydrodynamics and showed that
generally (when all baryon species are superfluid), 
it contains {\it sixteen} bulk viscosity coefficients.

({\it iii}) We have calculated and analyzed the sixteen bulk viscosity 
coefficients assuming they are generated by non-equilibrium 
reactions (\ref{s})--(\ref{ll}) 
of particle mutual transformations. 
We have shown that only {\it three} of them are independent. 
All other 13 bulk viscosities can be expressed through these three
using Eqs.\ (\ref{onsager11}) and (\ref{Sp})--(\ref{additional2}).
In addition, we have demonstrated that Eq.\ (\ref{xi22}) 
for the bulk viscosity coefficient $\xi_2$ 
formally coincides with Eq.\ (\ref{ksi})
for that in normal matter (however, the reaction rates 
$\lambda_l$ ($l=1,\ldots,4$) 
are affected by superfluidity).

Our results can be important for the studies of dynamical instabilities 
in pulsating superfluid neutron stars, 
especially for the studies of the r-mode instability. 
They can also be important for modeling of the thermal evolution 
of pulsating neutron stars 
and for analyzing rotochemical and gravitochemical heating 
of millisecond pulsars with superfluid nucleon-hyperon cores 
(for non-superfluid $npe$ matter of neutron stars,
this problem was considered 
in Refs.\ \cite{reisenegger95, fr05, rjfk06, jrf06}).

Let us notice, that to start such an analysis 
one needs to know the matrix $Y_{ik}$, which is 
the most important ingredient 
in the hydrodynamics of superfluid nucleon-hyperon mixture.
To our best knowledge, 
it has not been considered in the literature so far. 
We are planning to fill this gap 
and present its extensive calculations 
in a subsequent publication.

%Let us notice, that up to now it is impossible 
%to start such an analysis because
%the main ingredient of the hydrodynamics of superfluid 
%nucleon-hyperon mixtures, 
%the matrix $Y_{ik}$, is still unknown. 
%We are planning to calculate 
%this matrix in a subsequent paper.

%%%%%%%%%%%%%%%%%%%%%%%%%%%%%%%%%%%%%%%%%%%%%%%%%%%%%%%%%%
\section*{Acknowledgments}
%%%%%%%%%%%%%%%%%%%%%%%%%%%%%%%%%%%%%%%%%%%%%%%%%%%%%%%%%%
The authors are very grateful to
K.P. Levenfish and D.G. Yakovlev for allowing 
to use their code which calculates the third equation 
of state of Glendenning \cite{g85}.
The authors are also grateful 
to A.I. Chugunov for technical assistance
and to A.Y. Potekhin for critical comments.
This research was supported
by RFBR (grants 05-02-16245 and 05-02-22003)
and by the Federal Agency for Science and Innovations
(grants NSh 9879.2006.2 and NSh 2600.2008.2). 
One of the authors (M.E.G.) also acknowledges 
support from the Dynasty Foundation
and from the RF Presidential Program 
(grant MK-1326.2008.2).

%%%%%%%%%%%%%%%%%%%%%%%%%%%%%%%%%%%%%%%%%%%%%%%%%%%%%%%%%%

\end{document}